# Ferroelectric 2D Antimony Oxides with Wide Bandgaps

Romakanta Bhattarai[a], Kai Ni[b], and Xiao Shen[a*]

[a] Department of Physics and Materials Science, The University of Memphis, Memphis, TN, USA

[b] Department of Electrical and Microelectronic Engineering, Rochester Institute of Technology, Rochester, NY, USA

**Abstract**

The first two-dimensional (2D) polymorphs of antimony dioxide, namely, γ-$Sb_2O_4$ and δ-$Sb_2O_4$, are predicted using the evolutionary algorithm combined with first-principles density functional theory (DFT) calculations. Out-of-plane ferroelectricity is found in γ-$Sb_2O_4$, while in-plane ferroelectricity is found in δ-$Sb_2O_4$. The predicted dipole moment of γ-$Sb_2O_4$ and δ-$Sb_2O_4$ phases are 36.63 and 14.96 eÅ, respectively, implying that they can be good candidates for making ferroelectric devices. The calculations show that doping with other group V elements or applying strain can lower the switching energy barriers and thus facilitate switching. Results from GW calculations show indirect band gaps of 5.51 and 3.39 eV for γ-$Sb_2O_4$ and δ-$Sb_2O_4$ in their monolayers, respectively. Raman spectra are calculated to facilitate the experimental investigation of the predicted structures. The existence of both in-plane and out-of-plane 2D ferroelectricity and the large band gaps make this material system particularly interesting for potential applications.

**Introduction**

Ferroelectricity arises from a switchable spontaneous electric polarization in materials and is an intriguing feature in a wide range of materials. Ferroelectric materials have many potential applications, such as nonvolatile memory devices including ferroelectric capacitors and ferroelectric field effect transistor, and sensors.[1,2] One major challenge in using these materials in miniature devices is the depolarization field developed as the material's thickness decreases, which often destroys the ferroelectricity by inhibiting the electric polarization.[3,4] The two-dimensional (2D) van der Waals materials, however, offer an alternative path to overcome this problem, as recent findings of ferroelectric 2D materials make it possible to shrink the thicknesses of the ferroelectric devices down to atomic layers.[5–9] 2D ferroelectrics include CuCl,[10] MX (M=Sn,Ge;X=S,Se),[11] SnTe,[7] $In_2Se_3$,[6,12] $As_2X_3$ (X=S, Se, Te),[6,13] $WTe_2$,[9] $CuInP_2S_6$,[8] some of which are also realized in the experiment.[6–9,12,14,15] However, most of them show the in-plane electric polarization, which limits their applications in practical devices. The out-of-plane ferroelectricity, typically more desirable for nonvolatile memory applications, is extremely rare in 2D materials.[8,9,12] More importantly, a large bandgap is desirable for applications as it enables a large voltage being applied and low leakage current, but most 2D ferroelectrics known so far have small bandgaps. Under these circumstances, the Sb-O system is particularly interesting because of its rich structures, possibility to possess large band gaps, and interesting dielectric properties.[16–18]

---

[*] xshen1@memphis.edu

Several phases of antimony oxides have been investigated experimentally or theoretically. Among them, the α and β phases of $Sb_2O_3$ and $Sb_2O_4$ are the most studied. They are semiconductors with direct and indirect band gaps and smaller effective masses of the carriers.[19–23] These compounds have several applications ranging from catalysis to polymerization to coating.[24–29] Zhang et al. investigated a number of layered antimonene oxides theoretically and found their band gaps ranging from 0 to 2.28 eV, along with 18Sb-18O as a topological insulator.[30] Theoretical investigation on other 2D $Sb_2O_x$ (x= 1, 2, 3) was reported by Wolff et al.[16] Recently, Yang et al. reported experimental observation of new antimony oxide, $SbO_{1.93}$, in ultrathin samples, with an unusually large band gap (~ 6.3 eV) and a large static dielectric constant (~100).[18] Interestingly, antimony supphoiodide (SbSI) has long been known as a quasi-one-dimensional ferroelectric semiconductor,[31,32] which suggests the potential of ferroelectricity in low-dimensional Sb compounds.

In this paper, we propose two new layered phases of the antimony oxide, named $\gamma$-$Sb_2O_4$, and $\delta$-$Sb_2O_4$, from the evolutionary algorithm combined with the first-principles calculations. Out-of-plane ferroelectricity is predicted in $\gamma$-$Sb_2O_4$, while in-plane ferroelectricity is predicted in $\delta$-$Sb_2O_4$. We also show that the doping and strain can be used to adjust the switching energy barrier. The large dipole moment of $\gamma$-$Sb_2O_4$ and $\delta$-$Sb_2O_4$ imply that they can be good candidates for ferroelectric applications. Their monolayers feature large band gaps. The Raman properties are also predicted. The results highlight the rich structures and properties of the Sb-O system. The finding of out-of-plane and in-plane 2D ferroelectric oxides with large bandgaps are important for potential applications of the 2D ferroelectrics.

**Results**

We used a genetic algorithm in combination with the density functional theory (DFT) to search for the stable structures corresponding to the local energy minimum. Details of the computational methods can be found in Supplementary Information, Section I. A total of 13130 structures are generated from the genetic algorithm with varying unit sizes, stoichiometry, and dimensionality. Among them, we obtain a total of three stable phases of $Sb_2O_4$. One of them is the known α-$Sb_2O_4$, whose appearance validates our methodology. The other two phases of $Sb_2O_4$ have not been reported previously, and we name them $\gamma$- and $\delta$-$Sb_2O_4$. It is interesting that these new phases, which are the first 2D antimony oxides with the stoichiometric ratio of 2:4, are found by the genetic algorithm in both the 3D and 2D searches. Figure 1 shows the crystal structures of the predicted phases. For δ-Sb$_2$O$_4$, we explored the ABAB and ABBA types of stackings and found that the energy of ABAB stacking is lowered by 0.02 eV/atom. Therefore, we only focus on ABAB type of stackings, as shown in Fig. 1(c, d).

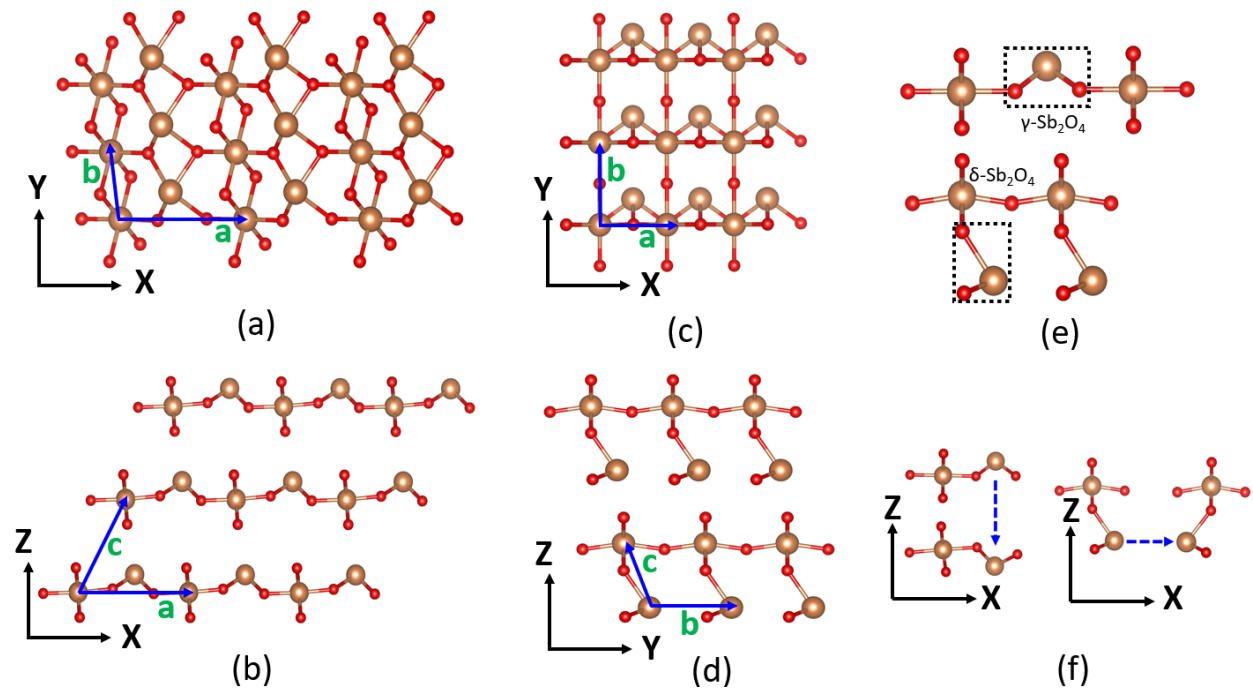

Table 1: Optimized lattice parameters of the γ-Sb$_2$O$_4$, δ-Sb$_2$O$_4$, and ε-Sb$_2$O$_4$

| Phase | a (Å) | b (Å) | c (Å) | α (°) | β (°) | γ (°) | Volume (Å$^3$) | Bravais Lattice |
|---|---|---|---|---|---|---|---|---|
| γ-Sb$_2$O$_4$ (bulk) | 6.55 | 3.06 | 5.87 | 75.4 | 61.4 | 90.2 | 99.09 | Triclinic |
| γ-Sb$_2$O$_4$ (monolayer) | 6.50 | 3.05 | - | 90 | 90 | 90 | - | Rectangular |
| δ-Sb$_2$O$_4$ (bulk) | 3.18 | 3.88 | 6.99 | 102.8 | 90 | 90 | 84.38 | Triclinic |
| δ-Sb$_2$O$_4$ (monolayer) | 3.16 | 3.88 | - | 90 | 90 | 90 | - | Rectangular |

First, we examine the bonding configurations in these Sb$_2$O$_4$ phases. In γ-Sb$_2$O$_4$, two types of covalent bonding between antimony and oxygen atoms are observed, as clearly shown in Fig. 1(a). Type one consists of an Sb atom bonded octahedrally (6-fold) with six neighboring oxygen atoms, whereas in type two, each Sb atom is tetrahedrally (4-fold) bonded to four oxygen atoms. Four among the six oxygen atoms around the octahedrally bonded Sb are two-fold coordinated, connecting only with the octahedral Sb atoms. On the other hand, all the four oxygens around the tetrahedrally bonded Sb are three-fold coordinated, with two bonds connecting to tetrahedrally bonded Sb atoms and the third bond connecting to octahedrally bonded Sb atoms. In δ-Sb$_2$O$_4$, a slight modification in bonding between Sb and O is observed. As shown in Fig. 1(b), one type of Sb is octahedrally (6-fold) bonded with six oxygen atoms as in the previous case, whereas the other Sb atom is bonded to only three oxygen atoms (3-fold). The oxygen atoms have two-fold and three-fold covalent bonds with Sb, similar to the case of γ-$Sb_2O_4$. Interestingly, in δ-$Sb_2O_4$, the two types of Sb-O bonds are in two different planes within a single layer, while in γ-$Sb_2O_4$, the two bonding types appear alternatingly along one direction within the same plane (Figure 1(e)). A detailed analysis of bonding environments among the γ- and δ-$Sb_2O_4$ phases and their corresponding bond lengths are presented in Supplementary Material, Section III.

These new layered $Sb_2O_4$ phases exhibit interesting ferroelectric properties, as the 4-fold and 3-fold Sb in γ- and δ-$Sb_2O_4$ are associated with electric dipoles switchable by an external electric field. As a result of their corresponding crystal structures, the γ-$Sb_2O_4$ phase exhibits an out-of-plane ferroelectric switching, and the δ-$Sb_2O_4$ phase exhibits an in-plane dipole switching (Fig. 1(f)). The calculated switching barriers for bulk phases of γ- and δ-$Sb_2O_4$ are shown in Figure 2. The maximum switching energy barrier for the dipole in γ-$Sb_2O_4$ is found to be 0.51 eV (Figure 2(a)), whereas, for δ-$Sb_2O_4$, it is 0.43 eV (Figure 2(b)). We also explore the feasibility of reducing the energy barrier through doping at the Sb site. It is found that the barrier is lowered to 0.44 eV upon doping of Bi in γ-$Sb_2O_4$, while it is decreased to 0.29 eV upon doping of As in δ-$Sb_2O_4$. These values are comparable to the switching barriers in AlN and 2D $In_2Se_3$[33,12], and smaller than $GaFeO_3$.[34] We also investigate the effect of strain in the γ-$Sb_2O_4$ phase by applying a tensile strain of +3% along the 'a' & 'b' axis and -3% along the 'c' axis. The energy barrier is decreased by 0.13 eV to 0.38 eV, suggesting that the strain can be another method to decrease the switching barrier in the γ-$Sb_2O_4$. The electric dipole moment in γ-$Sb_2O_4$ is 36.63 eÅ per unit cell in the out-of-plane direction, whereas in δ-$Sb_2O_4$, the moment is 14.96 eÅ per unit cell along the in-plane direction. These values are comparable with the croconic acid and $GaFeO_3$,[34,35] and imply that they can be good candidates for ferroelectric applications.

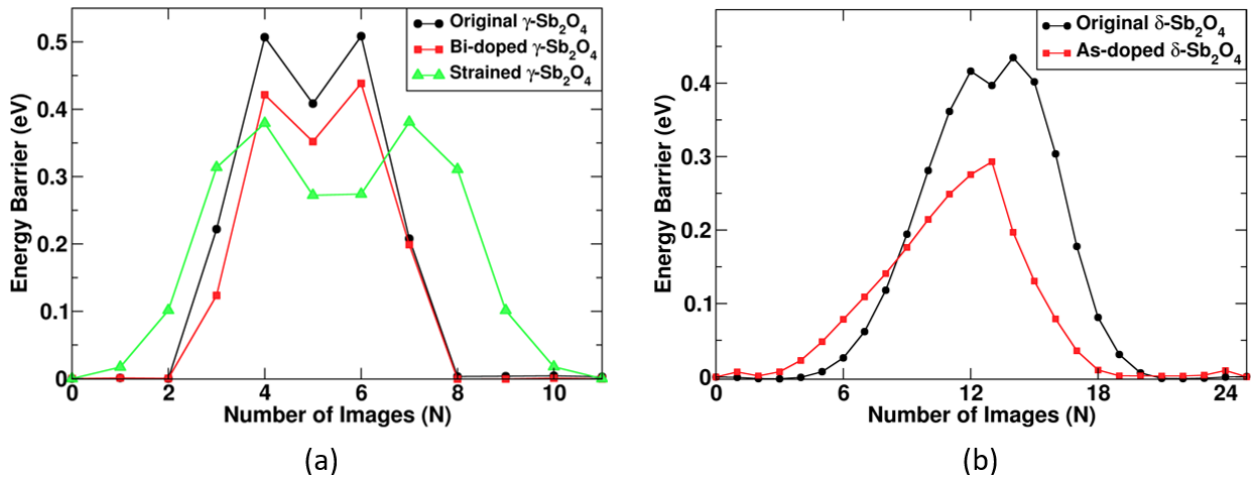

(a)      (b)

Figure 2: Switching energy barriers of (a) bulk γ-Sb$_2$O$_4$ and (b) bulk δ-Sb$_2$O$_4$ phases using the solid-state NEB method. The slight asymmetry is due to the influence of stacking.

The structural anisotropy in the Sb$_2$O$_4$ phases enables them to possess interesting properties, including ferroelectricity which arises from the breaking of the 180° inversion symmetry in-plane (δ phase) and out-of-the plane (γ phase). Further, the variability of Sb-O bonds arising from the lattice distortion results in the double- and single-well soft mode potentials in γ- and δ- phase, resembling some of the previously studied ferroelectrics[6,11,13]. In SbSI, a well-known Sb-based ferroelectric material investigated several decades ago, and the ferroelectric behavior is associated with thermal fluctuations, change in electronic structure, atomic disorder, and phonon interactions,[36] all closely related to the variability of bonding between Sb and S/I. The finding of 2D ferroelectric Sb$_2$O$_4$ is another manifestation of the rich behavior of Sb-based compounds and its potential for future discoveries. As we will discuss below, the Sb-O system further provides an avenue for ferroelectrics with large band gaps, which are beneficial for the applications.

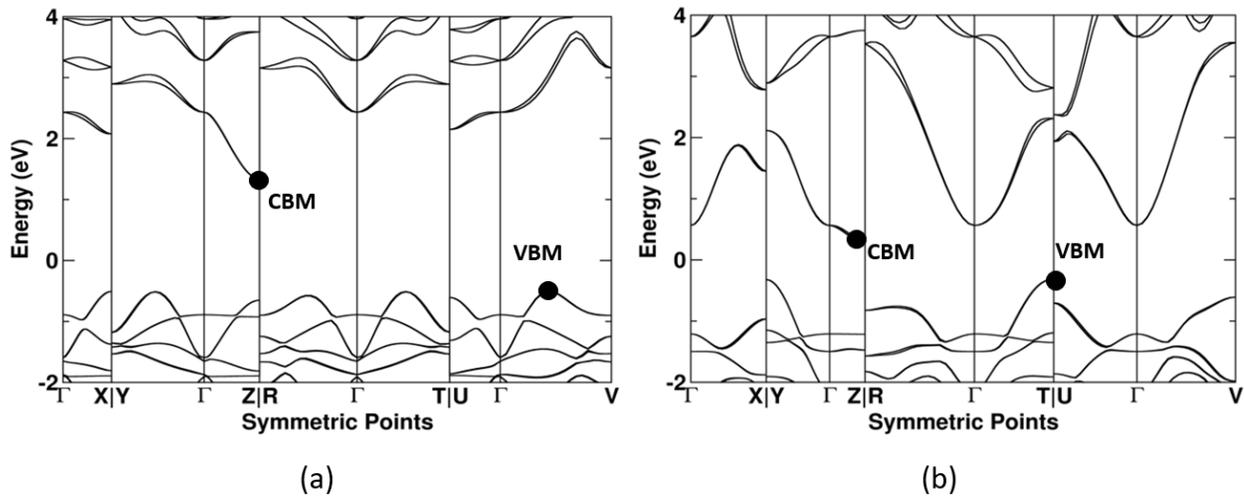

(a)  (b)

Figure 3: Band structures of (a) bulk γ-Sb$_2$O$_4$, and (b) bulk δ-Sb$_2$O$_4$ using DFT method considering the spin orbit interaction.

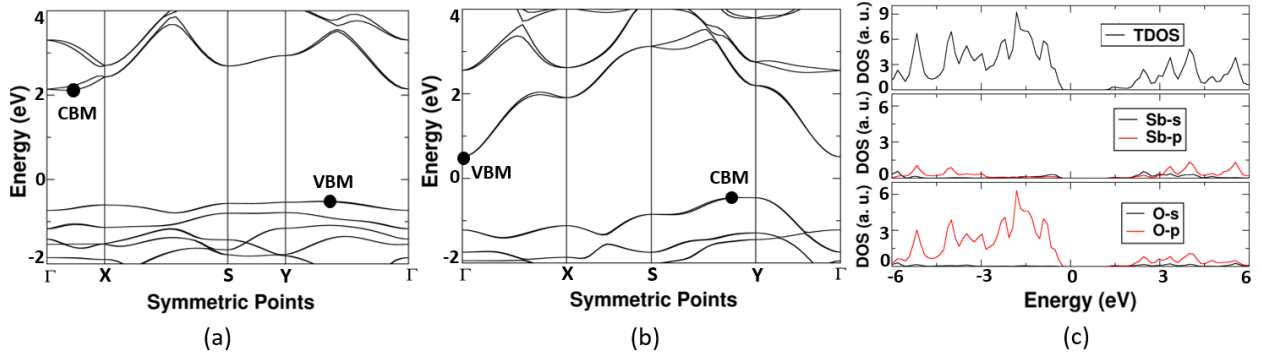

Figure 4: (a, b) DFT band structures of γ-Sb$_2$O$_4$, and δ-Sb$_2$O$_4$ monolayers (c) density of states of bulk γ-Sb$_2$O$_4$. Spin-orbit interaction is considered into account.

The electronic band structures of γ-Sb$_2$O$_4$ and δ-Sb$_2$O$_4$ using the standard DFT method are shown in Figure 3. The spin-orbit effects are considered in the calculations. In the bulk γ-Sb$_2$O$_4$ (Fig. 3(a)), the maximum of the valence band (VBM) lies between the Γ and V points, whereas the minimum of the conduction band (CBM) lies at the Z point in the Brillouin zone, giving rise to an indirect band gap of 1.88 eV. In bulk δ-Sb$_2$O$_4$ (Fig. 3(b)), the VBM lies between Γ and Z (very close to Z), and CBM lies at T, giving rise to an indirect band gap of 0.63 eV. The band structures of corresponding monolayers are shown in Figs. 4(a) and 4(b). In γ-Sb$_2$O$_4$ monolayer, VBM and CBM are located between Γ & X and Γ & Y, which give rise to an indirect band gap of 2.63 eV, whereas in δ-Sb$_2$O$_4$ monolayer, they are located at G, and between S & Y, thus making an indirect band gap of 0.96 eV. We also calculated the density of states (DOS) in each configuration. The compositions of DOS are similar in all cases, and therefore only the DOS of γ-Sb$_2$O$_4$ is presented in Figure 4(c). The p-orbitals of Sb and O atoms are dominant in both the valence band and conduction band (with p-orbitals of O in higher magnitude), while contributions from the respective s-orbitals are negligible.

Table 2: Band gaps of γ-Sb$_2$O$_4$, and δ-Sb$_2$O$_4$ phases under different methods

| Phase | DFT Band gap (eV) | HSE Band gap (eV) | GW Band gap (eV) |
| --- | --- | --- | --- |
| γ-Sb$_2$O$_4$ (bulk) | 1.84 | 2.99 | 3.66 |
| γ-Sb$_2$O$_4$ (monolayer) | 2.63 | 3.85 | 5.51 |
| δ-Sb$_2$O$_4$ (bulk) | 0.63 | 1.56 | 2.25 |
| δ-Sb$_2$O$_4$ (monolayer) | 0.96 | 2.05 | 3.39 |

To get further insight into the electronic properties, we perform the Bader charge analysis in the bulk phases. In γ-$Sb_2O_4$, each octahedrally bonded Sb atom loses a charge of 2.66e (e being the electron charge of -1.67x$10^{-19}$ C) to oxygen, whereas each tetrahedrally bonded Sb atom loses relatively less charge (1.83e). Each two-fold coordinated oxygen receives a charge of 1.11e, whereas each three-fold coordinated oxygen receives a slight more charge of 1.14e. In δ-$Sb_2O_4$, the corresponding values are 2.28e & 1.81e for Sb atoms, and 1.13e & 1.15e for the oxygen atoms. This charge transfer mechanism is consistent with the fact that the oxygen is more electronegative than the antimony atom.

As DFT is known to underestimate the band gap, we perform the hybrid DFT (HSE) calculation (including spin-orbit interaction) to better approximate the gap values.[37–39] We find the corresponding band gaps as 2.99 and 3.85 eV for the bulk and monolayer γ-$Sb_2O_4$, 2.05 and 1.56 eV for the bulk and monolayer δ-$Sb_2O_4$, respectively. To further improve the prediction of the band gaps of these phases, we employ the GW method[40–42] on top of DFT to include the quasiparticles effect. As the convergence under the GW method requires a sufficiently large number of empty conduction bands, we include 60 (17 of them are occupied) bands in our calculations, making sure that the supplied empty bands are sufficient for accurate results. From the GW approximation, the QP effect increases the band gaps to 3.66 eV, 5.55 eV, 2.25 eV, and 3.39 eV for bulk γ-$Sb_2O_4$, monolayer γ-$Sb_2O_4$, bulk δ-$Sb_2O_4$, and monolayer δ-$Sb_2O_4$, respectively (see Table 2). The large band gap values, especially in γ- and δ-$Sb_2O_4$ monolayers, are beneficial for their potential applications as a ferroelectric material in electronic devices.

Raman spectroscopy is a widely used technique to characterize materials. To facilitate the experimental exploration of the predicted $Sb_2O_4$ structures, we calculated the Raman spectra as shown in Figure 5 for the bulk and monolayers. In the bulk γ-$Sb_2O_4$ (Fig. 5(a)), two major Raman peaks are observed at 525 and 583 $cm^{-1}$, whereas in the monolayer (Fig. 5(b)), four peaks are observed at 444, 558, 610, and 699 $cm^{-1}$. Two major Raman peaks are observed at 511 and 635 $cm^{-1}$ in bulk δ-$Sb_2O_4$ (Fig. 5(c)) and 378 and 526 $cm^{-1}$ in the monolayer (Fig. 5(d)). In addition to the major peaks, some minor Raman peaks are also shown in the spectra. The detailed tabulation of the vibration modes is shown in Supplementary Information, Section IV. In addition to the Raman spectra, we also present the simulated XRD results using Cu-k alpha method in Supplementary Information, Section V to further assist experiments.

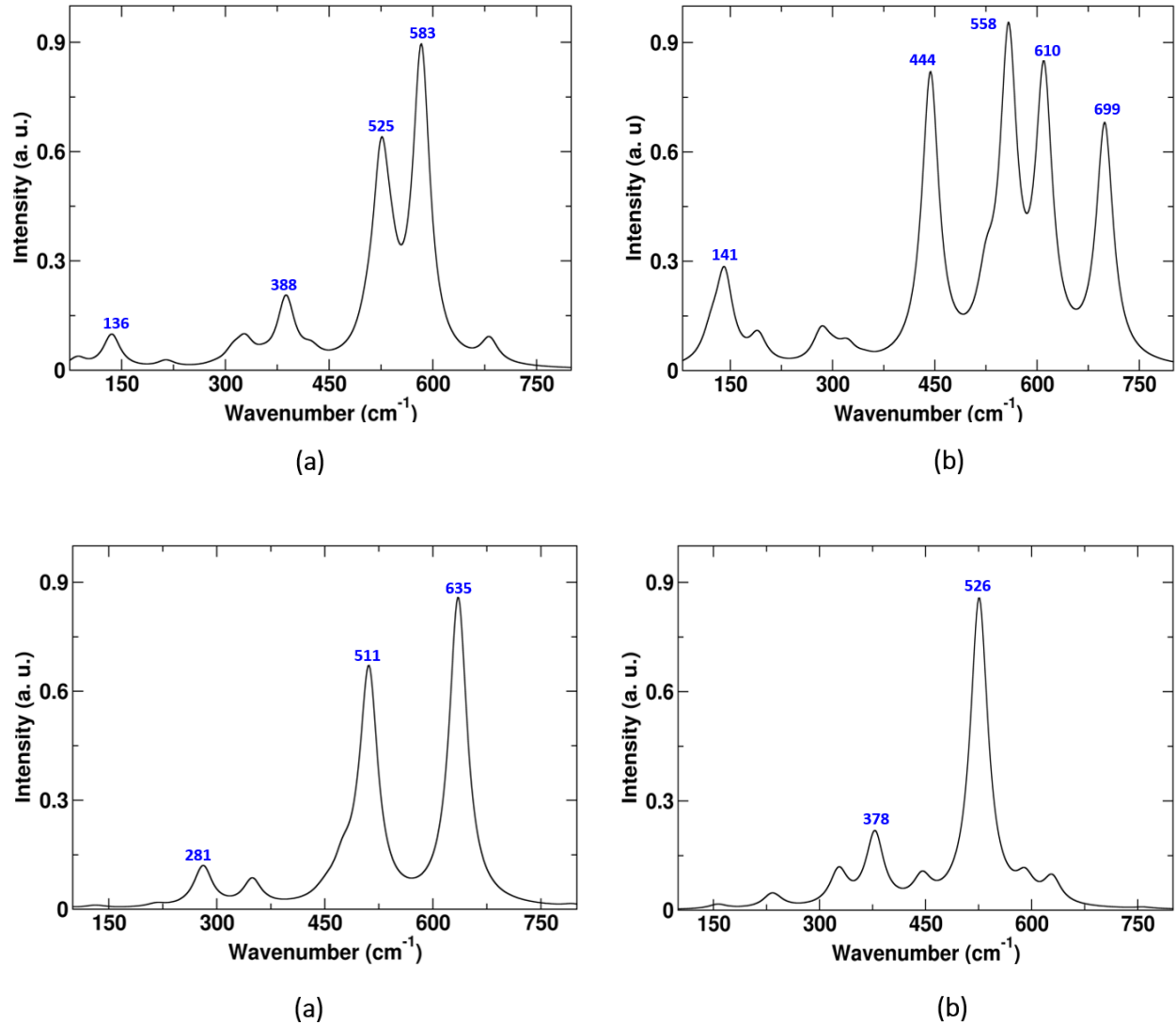

Figure 5: Calculated Raman spectra of (a) bulk γ-Sb$_2$O$_4$, (b) γ-Sb$_2$O$_4$ monolayer, (c) bulk δ-Sb$_2$O$_4$, and (d) δ-Sb$_2$O$_4$ monolayer.

**Conclusions**

Using the evolutionary algorithm combined with the first-principles DFT calculations, we predict the first two-dimensional (2D) polymorphs of antimony dioxide, namely, γ-Sb$_2$O$_4$ and δ-Sb$_2$O$_4$. Out-of-plane ferroelectricity is found in γ-Sb$_2$O$_4$, while in-plane ferroelectricity is found in δ-Sb$_2$O$_4$. We also show that doping with other group V elements or applying strain can lower the switching energy barriers. The dipole moment of γ-Sb$_2$O$_4$ and δ-Sb$_2$O$_4$ are calculated as 36.63 and 14.96 eÅ per unit cell, respectively, implying they are good candidates for ferroelectric applications. Furthermore, calculations show the indirect band gaps of 5.51 and 3.39 eV for γ- and δ-$Sb_2O_4$ in their monolayers, respectively. Raman spectra are predicted to pave the way for the experimental investigation of the predicted structures. The existence of both in-plane and out-of-plane 2D ferroelectricity and the large band gap are notable features of these new phases, which may be of interest for potential applications.

**Supplementary Material**

Computational details, stability of γ- and δ-$Sb_2O_4$ monolayers including interlayer coupling energy, phonon spectra and AIMD graphs, detailed bonding analysis, vibrational modes, and simulated XRD spectra.

**Acknowledgements**

This work at the University of Memphis was supported by the National Science Foundation under Grant No. DMR 1709528 and by the Ralph E. Powe Jr. Faculty Enhancement Awards from Oak Ridge Associated Universities (ORAU). Computational resources were provided by the High-Performance Computing Center (HPCC) at the University of Memphis. It is also partially supported by the U.S. Department of Energy, Office of Science, Office of Basic Energy Sciences Energy Frontier Research Centers program under Award Number DESC0021118.

**Competing Interests**

The authors declare no competing financial interests.

**Data Availability**

The data that support the findings of this study are available from the corresponding author upon reasonable request.

# Supplementary Material

## Ferroelectric 2D Antimony Oxides with Wide Bandgaps

Romakanta Bhattarai[a], Kai Ni[b], and Xiao Shen[a]*

[a]Department of Physics and Materials Science, The University of Memphis, Memphis, TN, USA

[b]Department of Electrical and Microelectronic Engineering, Rochester Institute of Technology, Rochester, NY, USA

*xshen1@memphis.edu

## I. Computational Details

The computational study in this work is carried out as follows. First, we use the genetic algorithm in combination with the density functional theory (DFT) to search for stable structures corresponding to the local energy minimum. Initially, a set of 30 structures are generated randomly from the space group set, and this number is preserved in each generation afterward. The calculation is stopped when the most stable structure remains unchanged in the fifteen generations. The best candidates in every generation are used to produce the offspring for the next generation. Heredity accounts for half of the candidates in each generation. The remaining half is produced from the space group symmetry (20%), soft mutation (10%), lattice mutation (10%), and permutation (10%). During the search, the number of atoms are varied from 3 to 30 with the stoichiometry of $Sb_2O_4$. The searches are carried out for both 3D and 2D dimensions. We use the USPEX[1,2] code for the genetic algorithm and the VASP[3] code for DFT calculations, with the Perdew-Burke-Ernzerhof (PBE)[4] type of exchange and correlation functional and the projected augmented wave (PAW)[5] type pseudopotential. The electronic and ionic convergence conditions are set to $10^{-5}$ and $10^{-4}$ eV, respectively, along with the plane-wave basis set having a maximum energy cutoff of 320 eV. The electric dipole moments in $\gamma$-$Sb_2O_4$ and $\delta$-$Sb_2O_4$ phases are calculated using the Berry-phase method.[6,7]

The most stable structures predicted from the USPEX are relaxed in a tighter convergence criterion. A plane-wave basis set with the energy cutoff of 500 eV, in addition to the thresholds of $10^{-9}$ eV and $10^{-4}$ eV/Å for the electronic energy and force convergence, are considered in these calculations. For the integration of the Brillouin zone, Γ-centered k-points grids of 15 × 15 × 15 and 15 × 15 × 1 are chosen for the bulk and the corresponding monolayers, respectively. The dynamical stability of the predicted structures is tested using their respective supercells implemented in the Phonopy package,[8] which works under the finite displacement method. Furthermore, the thermal stability is confirmed by performing the ab initio molecular dynamics simulations at 500 K for 30 ps.

The energy barrier for switching is calculated using the solid-state nudged elastic band (SS-NEB) method.[9] The convergence criterion is set such that the force between any two successive steps is less than 0.01 eV/Å. The Raman spectra of the predicted structures are calculated using the density functional perturbation theory as implemented in the Quantum Espresso package.[10] Norm-conserving pseudopotentials generated via Rappe-Rabe-Kaxiras-Joannopoulos (RRKJ)[11] scheme along with the PBE functional is used. The energy cutoff for the plane-wave basis set is 80 Rydberg. The total energy convergence criteria are set to $10^{-12}$, and $10^{-14}$ Ry, respectively, for the self-consistency and the phonon calculations.

## II. Stability of γ- and δ-Sb$_2$O$_4$ Monolayers

We construct the monolayers of γ- and δ-Sb$_2$O$_4$ and test the dynamical stabilities. First, we calculate the interlayer coupling energies and plot them as functions of interlayer distances, as shown in Fig. S1. The plots confirm that the monolayers are energetically stable and indicate the possibility of experimental synthesis of these monolayers.

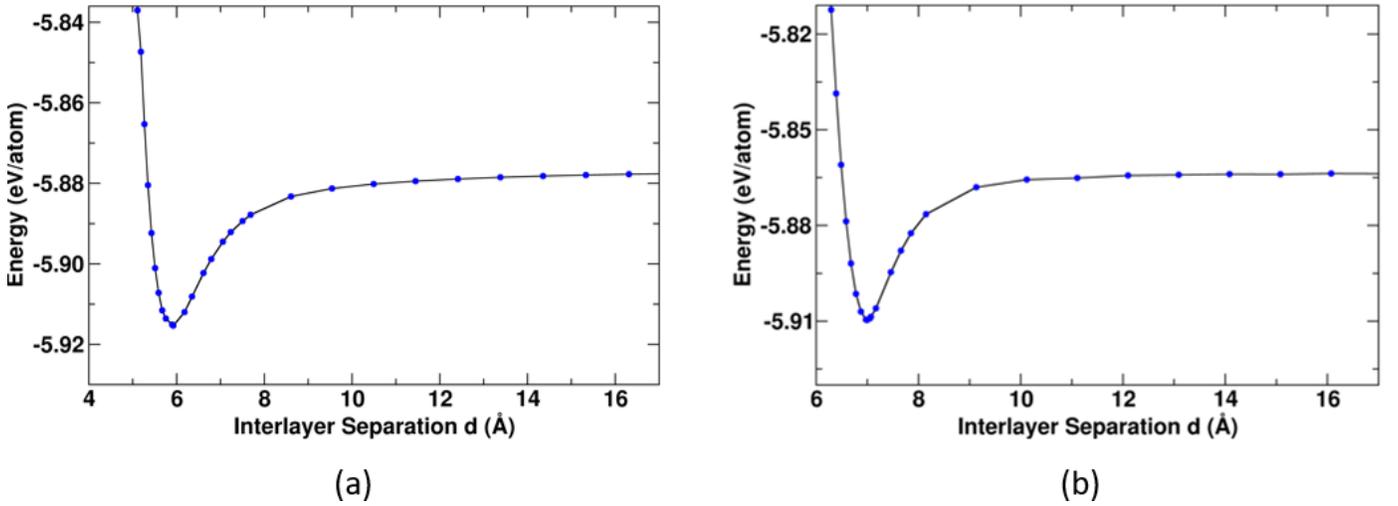

Figure S1: Coupling energy of as a function of interlayer separation (a) γ-Sb$_2$O$_4$, and (b) δ-Sb$_2$O$_4$

We then test the dynamical stability of all phases by calculating the phonon spectra using the Phonopy[8] package. Since we have confirmed that the monolayer of γ- and δ-Sb$_2$O$_4$ are energetically stable, testing the dynamical stability of these monolayers should be sufficient for the dynamical stability of the corresponding bulk phases. Therefore, we present the phonon dispersion spectra of γ- and δ-Sb$_2$O$_4$ monolayers in Figs. S2(a), and S3(a), respectively. There are no sizable imaginary frequencies in the phonon spectra in any case, which indicates that the corresponding structures are dynamically stable. A very small imaginary frequency appears near the Γ point in the γ-Sb$_2$O$_4$ phase, which is because of the numerical artifacts in the calculations. This could be removed by using larger supercell sizes or increasing the precisions (increased sampling points, larger energy cutoff, and tightening the criteria for electronic convergence).

In addition, we also test the thermal stability of the monolayers by performing the ab initio molecular dynamics (AIMD) simulation at an elevated temperature. The AIMD simulations are carried out in supercells of monolayers at 500 K for 30 ps with a time step of 2 fs. The potential energy profiles are plotted as a function of time, as shown in Figs. S2(b) and S3(b), for γ- and δ-$Sb_2O_4$ monolayers, respectively. Inserts represent the corresponding snapshots of the structures during the simulation. No sudden drop of the fluctuating potential energy, as well as no structure reconstruction throughout the simulation time, is observed, which implies the thermal stability of the structures.

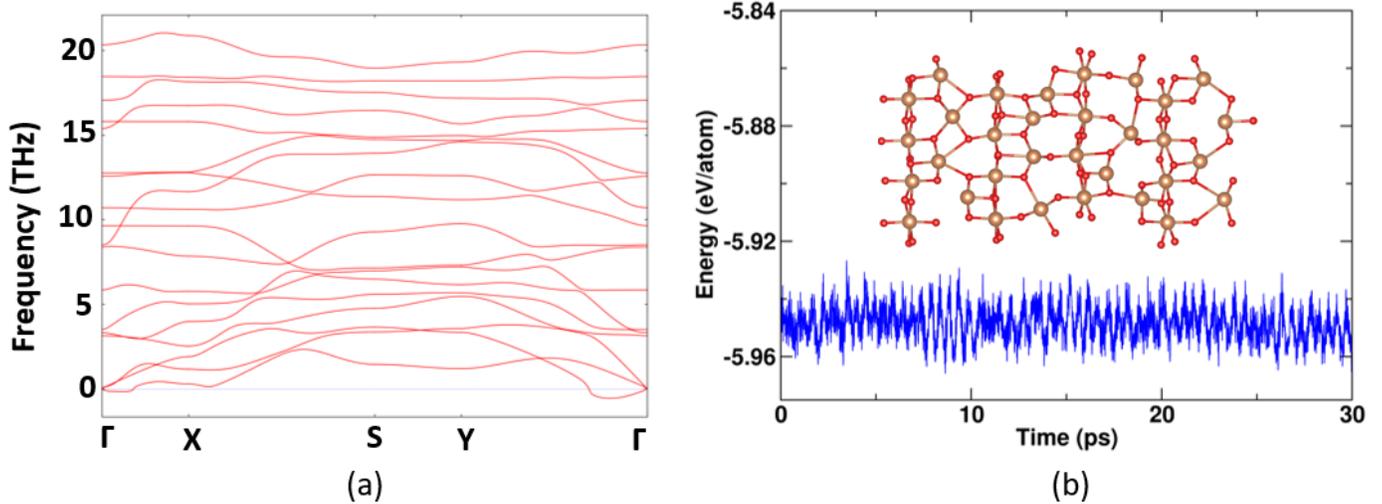

Figure S2: (a) Phonon band structures of monolayer γ-$Sb_2O_4$ (b) Potential energy profile of AIMD simulation of monolayer γ-$Sb_2O_4$ at 500 K for 30 ps. Insert shows a snapshot of structure during the simulation.

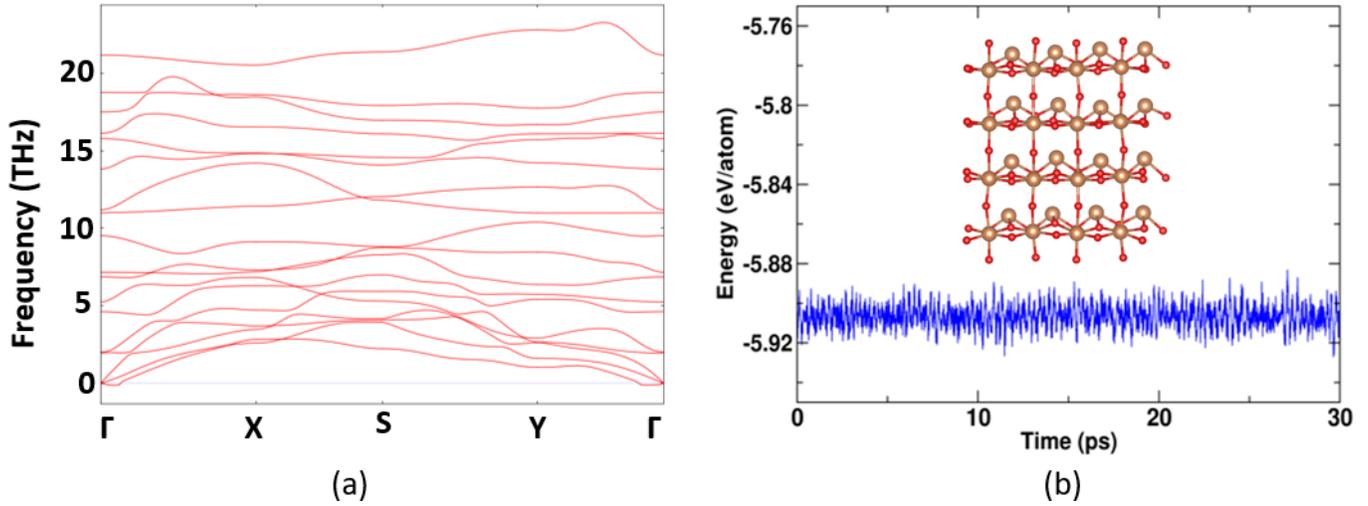

Figure S3: (a) Phonon band structures of monolayer δ-$Sb_2O_4$ (b) Potential energy profile of AIMD simulation of monolayer δ-$Sb_2O_4$ at 500 K for 30 ps. Insert shows a snapshot of structure during the simulation.

## III. Bonding Analysis on γ- and δ-$Sb_2O_4$

To get a deeper insight into the bonding nature among γ- and δ-$Sb_2O_4$, we plot each phase separately, highlighting every Sb-O bond in Fig. S4. Each ion is labeled to facilitate comparison with the bond lengths in Table S1.

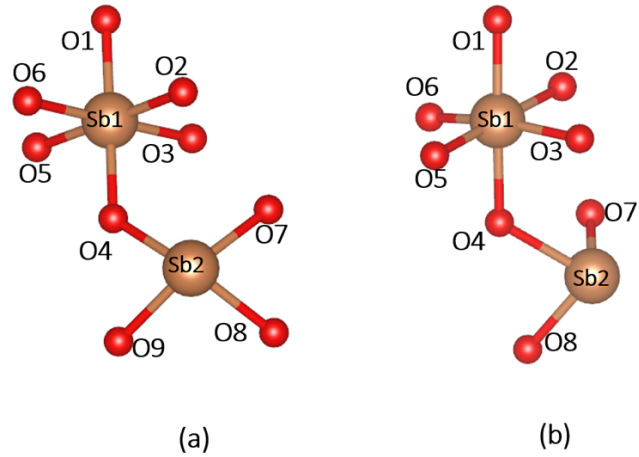

Figure S4: Bonding environments of Sb ions in (a) γ-$Sb_2O_4$, and (b) δ-$Sb_2O_4$

The structural configuration of γ-$Sb_2O_4$ looks similar to δ-$Sb_2O_4$; however, in reality, they differ in various aspects, including bond lengths, bond angles, et al., which give rise to separate properties between them. Table S1 below compares the bonding environments between the two $Sb_2O_4$ structures.

Table S1: Comparison of bonding environments among γ-$Sb_2O_4$, and δ-$Sb_2O_4$

| Phases | Bonding pairs | Bond lengths (Å) | Bonding pairs | Bond lengths (Å) |
|---|---|---|---|---|
| γ-$Sb_2O_4$ (bulk) | Sb1-O1 | 2.067 | Sb2-O4 | 2.153 |
| | Sb1-O2 | 2.012 | Sb2-O7 | 2.159 |
| | Sb1-O3 | 2.028 | Sb2-O8 | 2.148 |
| | Sb1-O4 | 2.054 | Sb2-O9 | 2.157 |
| | Sb1-O5 | 2.027 | | |
| | Sb1-O6 | 2.012 | | |
| γ-$Sb_2O_4$ (monolayer) | Sb1-O1 | 2.116 | Sb2-O4 | 2.159 |
| | Sb1-O2 | 2.038 | Sb2-O7 | 2.159 |
| | Sb1-O3 | 1.991 | Sb2-O8 | 2.128 |
| | Sb1-O4 | 2.055 | Sb2-O9 | 2.128 |
| | Sb1-O5 | 1.991 | | |
| | Sb1-O6 | 2.038 | | |
| δ-$Sb_2O_4$ (bulk) | Sb1-O1 | 2.051 | Sb2-O4 | 2.213 |
| | Sb1-O2 | 2.080 | Sb2-O7 | 2.019 |
| | Sb1-O3 | 1.966 | Sb2-O8 | 2.019 |
| | Sb1-O4 | 2.080 | | |
| | Sb1-O5 | 1.051 | | |
| | Sb1-O6 | 1.957 | | |
| δ-$Sb_2O_4$ (monolayer) | Sb1-O1 | 1.992 | Sb2-O4 | 2.070 |
| | Sb1-O2 | 2.137 | Sb2-O7 | 2.025 |
| | Sb1-O3 | 2.001 | Sb2-O8 | 2.025 |
| | Sb1-O4 | 2.138 | | |
| | Sb1-O5 | 1.992 | | |
| | Sb1-O6 | 1.993 | | |

## IV. Vibrational Modes

Both γ-$Sb_2O_4$ and δ-$Sb_2O_4$ have eighteen vibrational modes, as each unit cell contains six atoms in both phases. We tabulate the frequencies of these modes in Table S2, followed by plotting the vibrational eigenvectors in their bulk phases. We believe the results can be helpful for searching for these phases experimentally.

Table S2: Raman modes and corresponding vibrational frequencies of γ-$Sb_2O_4$, and δ-$Sb_2O_4$

| Mode | γ-$Sb_2O_4$ bulk | γ-$Sb_2O_4$ monolayer | δ-$Sb_2O_4$ bulk | δ-$Sb_2O_4$ monolayer |
|---|---|---|---|---|

|    | (cm⁻¹) | (cm⁻¹) | (cm⁻¹) | (cm⁻¹) |
|----|--------|--------|--------|--------|
| 1  | 0      | 0      | 0      | 0      |
| 2  | 0      | 0      | 0      | 0      |
| 3  | 0      | 0      | 0      | 0      |
| 4  | 42     | 107    | 87     | 63     |
| 5  | 86     | 122    | 96     | 67     |
| 6  | 136    | 141    | 128    | 155    |
| 7  | 214    | 191    | 137    | 162    |
| 8  | 284    | 284    | 216    | 232    |
| 9  | 311    | 301    | 273    | 240    |
| 10 | 328    | 321    | 281    | 327    |
| 11 | 354    | 346    | 349    | 370    |
| 12 | 388    | 427    | 442    | 378    |
| 13 | 424    | 444    | 454    | 445    |
| 14 | 505    | 525    | 475    | 526    |
| 15 | 525    | 525    | 495    | 542    |
| 16 | 539    | 558    | 511    | 591    |
| 17 | 583    | 610    | 635    | 629    |
| 18 | 682    | 699    | 794    | 758    |

The first three modes in each phase having zero vibrational frequency represent the collective translational motion of the crystal units in the same direction (see below). The corresponding vibrational modes are presented in Figures S5 and S6. The antimony and oxygen atoms are represented by gray and red colors, respectively. The directions of displacements are shown by green arrows.

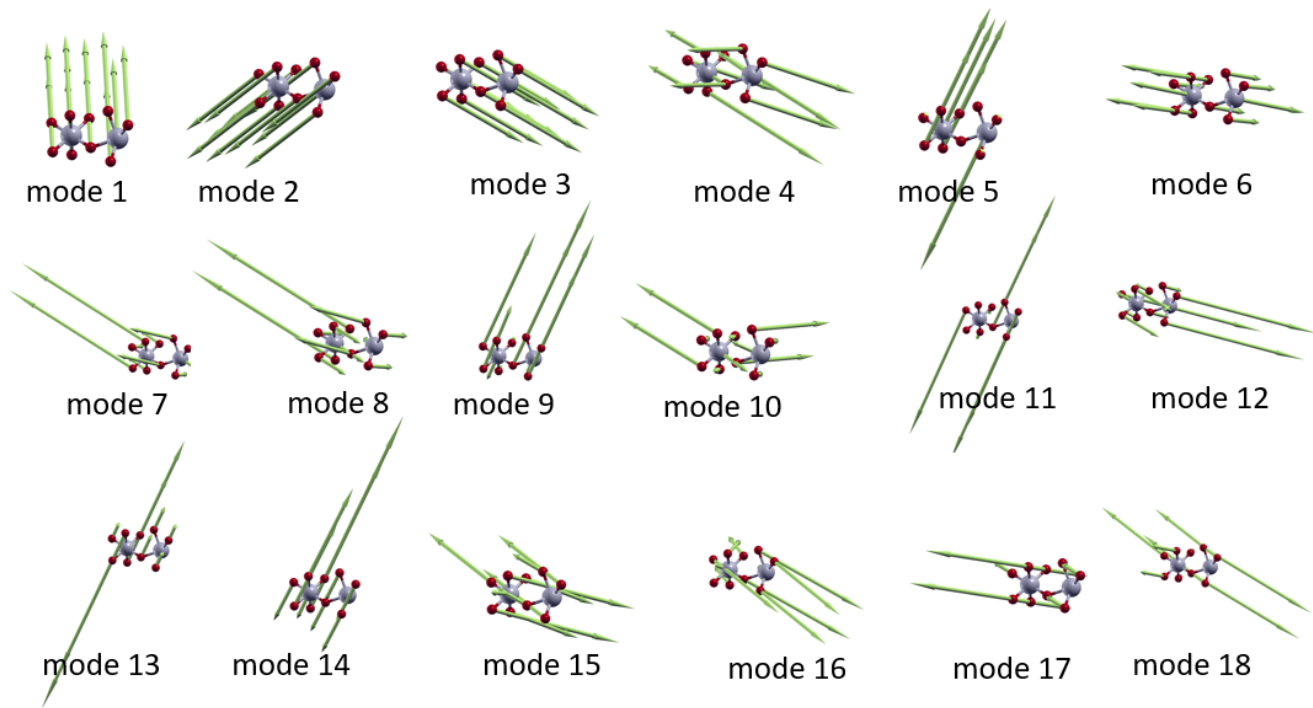

Figure S5: Eighteen vibrational modes of bulk γ-Sb$_2$O$_4$.

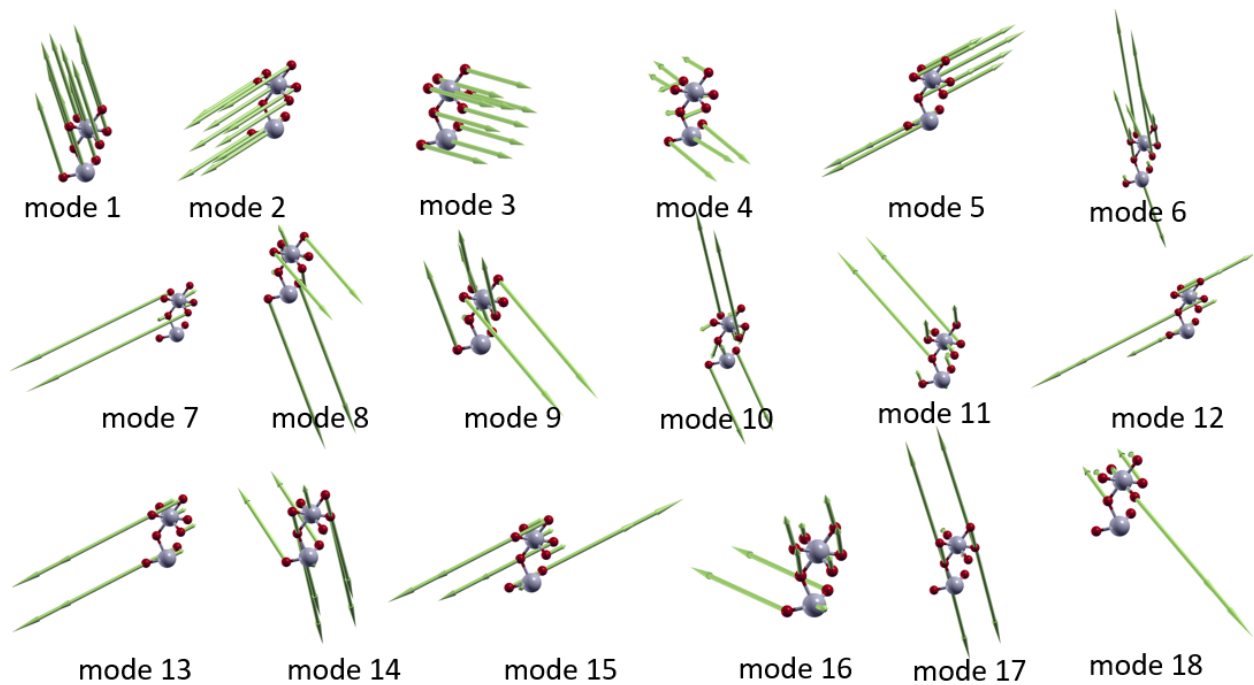

Figure S6: Eighteen vibrational modes of bulk δ-Sb$_2$O$_4$.

## V. X-Ray Diffraction (XRD) Spectra

In addition to Raman spectra, we also present the simulated XRD results using Cu-k alpha method in Table S3 below to further help the experimental investigations.

Table S3: Simulated XRD results of bulk $\gamma$-$Sb_2O_4$, and $\delta$-$Sb_2O_4$ using Cu-k alpha method

| S. N. | $\gamma$-$Sb_2O_4$ $2\theta$ (°) | $\delta$-$Sb_2O_4$ $2\theta$ (°) |
|---|---|---|
| 1 | 15.852 | 23.499 |
| 2 | 16.571 | 24.179 |
| 3 | 18.160 | 26.136 |
| 4 | 27.040 | 29.456 |
| 5 | 30.019 | 30.951 |
| 6 | 31.202 | 36.848 |
| 7 | 32.171 | 37.303 |
| 8 | 33.501 | 38.655 |
| 9 | 36.839 | 39.351 |
| 10 | 37.499 | 41.070 |
| 11 | 40.404 | 46.826 |
| 12 | 47.924 | 49.224 |
| 13 | 48.491 | 50.768 |
| 14 | 49.261 | 53.763 |
| 15 | 50.087 | 55.432 |
| 16 | 51.112 | 56.536 |
| 17 | 53.938 | 57.813 |
| 18 | 55.800 | 59.478 |
| 19 | 60.314 | 63.516 |
| 20 | 62.029 | 64.505 |
| 21 | 67.293 | 68.524 |

The XRD data are then plotted by showing the corresponding miller indices (hkl) in Figure S7 below.

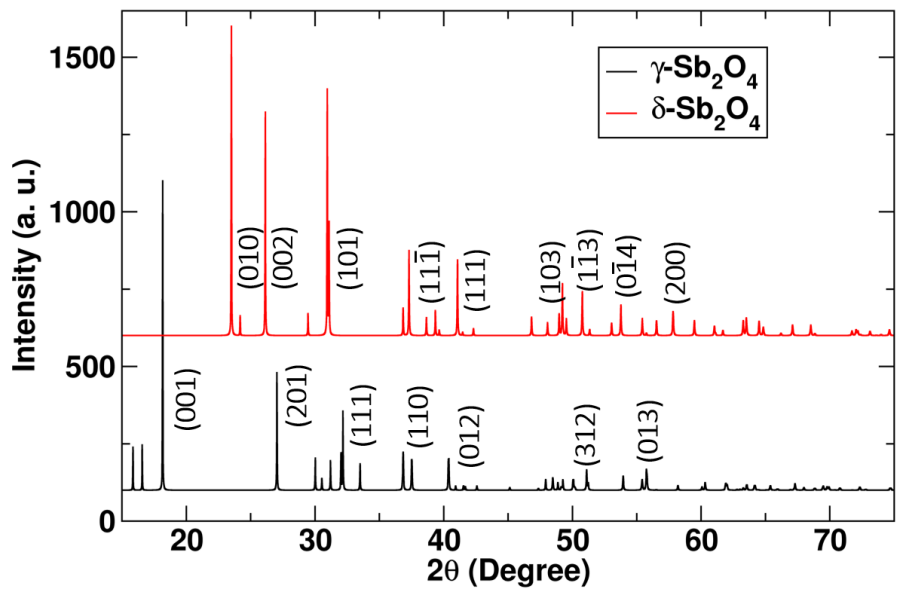

Figure S7: Plot of simulated XRD spectra of bulk γ-$Sb_2O_4$, and δ-$Sb_2O_4$ using Cu-k alpha method